\documentstyle[12pt]{article}

\newcommand{\bd}{\begin{document}}
\newcommand{\ed}{\end{document}}
\newcommand{\bc}{\begin{center}}
\newcommand{\ec}{\end{center}}
\newcommand{\be}{\begin{eqnarray}}
\newcommand{\ee}{\end{eqnarray}}
\renewcommand{\thefootnote}{\alph{footnote}}
\newcommand{\se}{\section}
\newcommand{\sse}{\subsection}
\newcommand{\bi}{\bibitem}
\def\figcap{\section*{Figure Captions\markboth
     {FIGURECAPTIONS}{FIGURECAPTIONS}}\list
     {Figure \arabic{enumi}:\hfill}{\settowidth\labelwidth{Figure 999:}
     \leftmargin\labelwidth
     \advance\leftmargin\labelsep\usecounter{enumi}}}
\let\endfigcap\endlist \relax
\begin{document}

\begin{titlepage}

 \vskip 0.5in
 \null
\begin{center}
 \vspace{.15in}
{\LARGE {\bf
Rare $\Lambda _{b}\rightarrow \Lambda $ l$^{+}$ l$^{-}$ Decays with
Polarized $\Lambda $ }
}\\
\vspace{1.0cm}  \par
 \vskip 2.1em
 {\large
  \begin{tabular}[t]{c}
{\bf Chuan-Hung Chen$^a$ and C.~Q.~Geng$^b$}
\\
\\
       {\sl ${}^a$Department of Physics, National Cheng Kung University}
\\   {\sl  $\ $Tainan, Taiwan,  Republic of China }
\\
\\
{\sl ${}^b$Department of Physics, National Tsing Hua University}
\\  {\sl  $\ $ Hsinchu, Taiwan, Republic of China }
\\
   \end{tabular}}
 \par \vskip 5.3em

 {\Large\bf Abstract}
\end{center}
We investigate the rare baryonic exclusive decays of $\Lambda_b\to
\Lambda l^+ l^-\ (l=e,\mu,\tau )$ with polarized $\Lambda $. Under
the approximation of the heavy quark effective theory (HQET), in
the standard model we derive the differential decay rates and
various polarization asymmetries by including lepton mass effects.
We find that with the long-distance effects the decay branching
ratios are $5.3\times 10^{-5}$ for $\Lambda_b\to \Lambda l^+ l^-\
(l=e,\mu)$ and  $1.1\times 10^{-5}$ for $\Lambda_b\to \Lambda
\tau^+\tau^-$. The effects of new physics in the decay rates are
also discussed. The integrated  longitudinal $\Lambda$
polarizations
are $-0.31$ and $-0.12$, while that of the normal ones $0.02$ and
$0.01$, for di-muon and tau modes, respectively. The CP-odd
transverse polarization of $\Lambda$ is zero in the standard model
but it is expected to be sizable in some theories with new physics.


\end{titlepage}

\section{ Introduction}

It is known that the recent interest in flavor physics has been focused in
the rare decays related to $b\to s l^+l^-$ induced by the flavor changing
neutral current (FCNC) due to the CLEO measurement of the radiative $b \to
s\gamma$ decay \cite{cleo}. In the standard model, these rare decays occur
at loop level and provide us information on the parameters of the
Cabibbo-Kobayashi-Maskawa (CKM) matrix elements \cite{ckm} as well as
various hadronic form factors. In the literature, most of studies have been
concentrated on the corresponding exclusive rare B meson decays such as $%
B\to K^{(*)}l^+l^-$ \cite{Bmeson}.

In this paper, we investigate the baryon decays of $\Lambda_b\to\Lambda
l^+l^-$ with $\Lambda $ being polarized. Unlike mesonic decays, the baryonic
decays could maintain the helicity structure of interactions in transition
matrix elements. Through this property, we will show that the polarization
asymmetries of $\Lambda$ are sensitive to right-handed couplings which are
suppressed in the standard model. Thus, these baryonic decays could be used
to search for physics beyond the standard model.

To study the exclusive bayonic decays, one of the most difficulties is to
evaluate the hadronic matrix elements. It is known that there are many form
factors for the matrix elements of $\Lambda_b$ to $\Lambda$, which are hard
to be calculated since they are related to the non-perturbative effect of
QCD. However, in heavy particle decays, HQET could reduce the number of form
factors and supply the information with respect to their relative size. In
our numerical calculations, we shall use the results in HQET. It is also
know that a large theoretical uncertainty in our calculation to the decays
arises from the long-distance (LD) effect. To reduce the uncertainty, we
shall study various kinematic regions to distinguish the LD contributions.
In our calculations, as a completeness, we will include the lepton mass,
which is important for the tau lepton mode.

The paper is organized as follows. In Sec.~2, we study the effective
Hamiltonian for the di-lepton decays of $\Lambda_b \to \Lambda l \bar{l} $
and form factors in the $\Lambda_b \to \Lambda$ transition. In Sec.~3, we
derive the general forms of the differential decay rates and the $\Lambda$
polarizations. In Sec.~4, we give the numerical analysis. We present our
conclusions in Sec.~5.

\section{Effective Hamiltonian and Form factors}

The effective Hamiltonian for the inclusive decay of $b\rightarrow
sl^{+}l^{-}$ is given by

\begin{equation}
{\cal H}=-4\frac{G_{F}}{\sqrt{2}}V_{tb}V_{ts}^{*}\sum_{i=1}^{10}C_{i}\left(
\mu \right) O_{i}\left( \mu \right) \,,  \label{Ham}
\end{equation}
where the expressions for the renormalized Wilson coefficients $C_{i}\left(
\mu \right) $ and operators $O_{i}\left( \mu \right) $ can be found in Ref.
\cite{Buras}. In terms of the Hamiltonian in Eq. (\ref{Ham}), the decay
amplitude is written as
\begin{eqnarray}
{\cal M} &=&\frac{G_{F}\alpha _{em}}{\sqrt{2}\pi }V_{tb}V_{ts}^{*}\left[
\bar{s}\left( C_{9}^{eff}\left( \mu \right) \gamma _{\mu }P_{L}-\frac{2m_{b}%
}{q^{2}}C_{7}\left( \mu \right) i\sigma _{\mu \nu }q^{\nu }P_{R}\right) b\;%
\bar{l}\gamma ^{\mu }l\right.  \nonumber \\
&&\left. +\bar{s}C_{10}\gamma _{\mu }P_{L}b\;\bar{l}\gamma ^{\mu }\gamma
_{5}l\right]  \label{Am}
\end{eqnarray}
with $P_{L(R)}=(1\mp \gamma _{5})/2$. We note that in Eq. (\ref{Am}), only
the term associated with Wilson coefficient $C_{10}$ is independent of the $%
\mu $ scale. We also note that the dominant contribution to the decay rate
is from the long-distance (LD) such as the c\={c} resonant states of $\Psi
,\Psi ^{\prime }...etc.$ To find out the LD effects for the B-meson decays,
in the literature \cite{DTP,LMS,AMM,OT,KS,Geng1}, both the factorization
assumption (FA) and the vector meson dominance (VMD) approximation have been
used. For the LD contributions in baryonic decays, we assume that the
parametrization is the same as that in the B meson decays. Hence, we include
the resonant effects (RE) by absorbing it to the corresponding Wilson
coefficients. The effective Wilson coefficient of $C_{9}^{eff}$ has the
standard form
\begin{equation}
C_{9}^{eff}=C_{9}\left( \mu \right) +\left( 3C_{1}\left( \mu \right)
+C_{2}\left( \mu \right) \right) \left( h\left( x,s\right) +\frac{3}{\alpha
_{em}^{2}}\sum_{j=\Psi ,\Psi ^{\prime }}k_{j}\frac{\pi \Gamma \left(
j\rightarrow l^{+}l^{-}\right) M_{j}}{q^{2}-M_{j}^{2}+iM_{j}\Gamma _{j}}%
\right) \,,  \label{effc9}
\end{equation}
where $h(x,s)$ describes the one-loop matrix elements of operators $O_{1}=%
\bar{s}_{\alpha }\gamma ^{\mu }P_{L}b_{\beta }\ \bar{c}_{\beta
}\gamma _{\mu }P_{L}c_{\alpha }$ and $O_{2}=\bar{s}\gamma ^{\mu
}P_{L}b\ \bar{c}\gamma _{\mu }P_{L}c$ as shown in Ref.
\cite{Buras}, $M_{j}$ and $\Gamma _{j}$ are the masses and widths
of intermediate states, and the factors $k_{j}$ are
phenomenological parameters for compensating the approximations of
FA and VMD and reproducing the correct branching ratios of
$B(\Lambda _{b}\to \Lambda J/\Psi \to \Lambda
l^{+}l^{-})=B(\Lambda_b\to \Lambda J/\Psi )\times B(J/\Psi \to
l^{+}l^{-})$ when we study the $\Lambda_b$ decays. We note that
by taking $k_{\Psi }\simeq -1/(3C_{1}+C_{2})$ and $B(\Lambda_b\to
\Lambda J/\Psi )=\left( 4.7\pm 2.8\right) \times 10^{-4}$, the
$k_{j}$ factors in the $\Lambda_b$ case are almost the same as
that in the B meson one. In this paper we take the Wilson
coefficients at the scale of $\mu \sim m_{b}\sim
5.0 $ GeV and their values are $C_{1}\left( m_{b}\right) =-0.226,$ $%
C_{2}\left( m_{b}\right) =1.096,$ $C_{7}\left( m_{b}\right) =-0.305,$ $%
C_{9}\left( m_{b}\right) =4.186,$ and $C_{10}\left( m_{b}\right) =-4.599$,
respectively.

It is clear that one of the main theoretical uncertainties in studying
exclusive decays arises from the calculation of form factors. In general
there are many form factors in exclusive baryon decays. However, the number
of the form factors can be reduced by the heavy quark effective theory
(HQET). With HQET, the hadronic matrix elements for the heavy baryon decays
could be parametrized as follows \cite{MR}
\begin{equation}
<\Lambda (p,s)\ |\ \bar{s}\ \Gamma \ b\ |\ \Lambda _{b}(\upsilon ,s^{\prime
})>=\bar{u}_{\Lambda }(p,s)\ \left\{ F_{1}(p\cdot v)+\not{v}\ F_{2}(p\cdot
v)\right\} \ \Gamma \ u_{\Lambda _{b}}(v,s^{\prime })  \label{hq}
\end{equation}
with $R=F_{2}\left( p.v\right) /F_{1}\left( p.v\right) $ where $v$ is the
four-velocity of heavy baryon and $\Gamma $ denotes the possible Dirac
matrix. 
Note that in terms of HQET there are only two independent form factors in
Eq. (\ref{hq}) for each $\Gamma $. In the following, we shall adopt the HQET
approximation to analyze the behavior of $\Lambda _{b}\to \Lambda l^{+}l^{-}$%
.

\section{ Differential Decay Rate and Polarizations}

In this section we present the formulas for the differential decay rates and
the longitudinal and normal $\Lambda$ polarizations of $\Lambda_b(p_{%
\Lambda_b})\to \Lambda (p_{\Lambda},s) l^{+}(p_{l^+})l^{-}(p_{l^-})$. In our
calculations, we have included the lepton masses. To study the $\Lambda $
spin polarization, we write the $\Lambda $ four-spin vector in terms of a
unit vector, $\hat{\xi}$, along the $\Lambda $ spin in its rest frame, as
\begin{eqnarray}
s_{0}\,=\,\frac{\vec{p}_{\Lambda }\cdot \hat{\xi}}{M_{\Lambda }},\qquad \vec{%
s}\,=\,\hat{\xi}+\frac{s_{0}}{E_{\Lambda }+M_{\Lambda }}\vec{p}_{\Lambda },
\end{eqnarray}
and choose the unit vectors along the longitudinal, normal, transverse
components of the $\Lambda $ polarization, to be
\begin{eqnarray}
\hat{e}_{L} &=&\frac{\vec{p}_{\Lambda }}{\left| \vec{p}_{\Lambda }\right| },
\nonumber \\
\hat{e}_{N} &=&\frac{\vec{p}_{\Lambda }\times \left( \vec{p}_{l^{-}}\times
\vec{p}_{\Lambda }\right) }{\left| \vec{p}_{\Lambda }\times \left( \vec{p}%
_{l^{-}}\times \vec{p}_{\Lambda }\right) \right| },  \nonumber \\
\hat{e}_{T} &=&\frac{\vec{p}_{l^{-}}\times \vec{p}_{\Lambda }}{\left| \vec{p}%
_{l^{-}}\times \vec{p}_{\Lambda }\right| }\,,  \label{uv}
\end{eqnarray}
respectively. The partial decay width for $\Lambda_b\rightarrow \Lambda \
l^{+}\ l^{-}\ ( l=e$ or $\mu$ or $\tau)$ is given by
\begin{eqnarray}
d\Gamma &=&\frac{1}{4M_{\Lambda _{b}}}\left| {\cal M}\right| ^{2}\left( 2\pi
\right) ^{4}\delta \left( p_{\Lambda _{b}}-p_{\Lambda
}-p_{l^+}-p_{l^-}\right)  \nonumber \\
&&\times \frac{d\vec{p}_{\Lambda }}{\left( 2\pi \right) ^{3}2E_{\Lambda }}
\frac{d\vec{p}_{l^+}}{\left( 2\pi \right) ^{3}2E_{1}}\frac{d\vec{p}_{l^-}}{%
\left( 2\pi \right) ^{3}2E_{2}}  \label{Dr0}
\end{eqnarray}
with
\begin{eqnarray}
|{\cal M}| ^{2}&=&\frac{1}{2}\left| {\cal M}^{0}\right| ^{2}\left[ 1+\left(
P_L\hat{e}_L+P_N\hat{e}_N+ P_T\hat{e}_T\right) \cdot \hat{\xi}\right] ,
\label{M1}
\end{eqnarray}
where $|{\cal M}^0| ^{2}$ is related to the decay rate for the unpolarized $%
\Lambda $ and $P_i\ (i=L,N,T)$ denote the longitudinal, normal and
transverse polarizations of $\Lambda$, respectively. Introducing
dimensionless variables of $\lambda _{t}=V_{tb}V_{ts}^{*}$, $\hat{t}%
=E_{\Lambda}/M_{\Lambda_{b}}$, $r=M_{\Lambda }^{2}/M_{\Lambda_{b}}^{2}$, $%
\hat{m}_l=m_{l}/M_{\Lambda _{b}}$, $\hat{m}_{b}=m_{b}/M_{\Lambda _{b}}$ and $%
\hat{s}=1+r-2 \hat{t}$, and integrating the angle dependence of the lepton,
the differential decay width in Eq. (\ref{Dr0}) can be rewritten as
\begin{eqnarray}
d\Gamma &=&\frac{1}{2}d\Gamma^0\left[ 1+\vec{P}\cdot \hat{\xi}\right]
\nonumber \\
d\Gamma^0 &=&\frac{G_F^2\alpha_{em}^2|\lambda_t|^2}{192\pi^5}
M_{\Lambda_b}^5 \sqrt{(\hat{t}^2-r)\left(1-\frac{4\hat{m}_l^2}{\hat{s}}%
\right)} \rho_0( \hat{t}) d\cos \theta _{\Lambda }d\hat{t},  \label{diffrate}
\end{eqnarray}
with
\begin{eqnarray}
\vec{P}&=& P_L\hat{e}_L+P_N\hat{e}_N+ P_T\hat{e}_T
\end{eqnarray}
and
\begin{eqnarray}
\rho_{0}\left( \hat{t}\right) &=& (\Gamma _{1}+\Gamma
_{2}+\Gamma_{3}+\Gamma_{4})
\end{eqnarray}
where 
\begin{eqnarray}
\Gamma_{1} &=&4\frac{\hat{m}_{b}^{2}}{\widehat{s}}|C_{7}|^{2} \left\{
-\left( F_1^2-F_2^2\right) \left( \hat{s}\ \hat{t}-4(1-\hat{t})(\hat{t}%
-r)\right) \right.  \nonumber \\
&& -2F_2(F_1\sqrt{r}+F_2\hat{t})\ \left( \hat{s}\ -4(1-\hat{t})^{2}\right) +8%
\frac{\hat{m}_l^2}{\hat{s}} \left( (F_{1}^{2}-F_{2}^{2})(1-\hat{t})(\hat{t}%
-r)\right.  \nonumber \\
&& \left. \left. +2F_2(F_1\sqrt{r}+F_2\hat{t})(1-\hat{t})^2 \right) -2\hat{m}%
_l^2 \left( (F_1^2+F_2^2)\ \hat{t}+2F_{1}F_{2}\sqrt{r}\right) \right\} ,
\nonumber \\
\Gamma_2 &=&12\hat{m}_{b}\mathop{\rm Re}C_{9}^{eff}C_{7}^{*}\left( 1+2\frac{%
\hat{m}_{l}^{2}}{\hat{s}}\right) \left[ (F_{1}^{2}-F_{2}^{2})(\hat{t}%
-r)+2F_{2}(F_{1}\sqrt{r}+F_{2}\hat{t})(1-\hat{t})\right] ,  \nonumber \\
\Gamma _3 &=& \left( |C_{9}^{eff}|^{2}+|C_{10}|^{2}\right) \left\{ \left(1-4%
\frac{\hat{m}_{l}^{2}}{\hat{s}}\right) \hat{s} \left[(F_{1}^{2}+F_{2}^{2})\
\hat{t}+2F_{1}F_{2}\sqrt{r}\right] \right.  \nonumber \\
&&+2(1+2\frac{\hat{m}_l^2}{\hat{s}}) \left( 1-\hat{t}\right) \left.\left[
\left( \hat{t}-r\right) (F_{1}^{2}-F_{2}^{2})+2F_{2}(F_{1}\sqrt{r}+F_{2}\hat{%
t}) \left( 1-\hat{t}\right) \right] \right\},  \nonumber \\
\Gamma _{4} &=&6\hat{m}_{l}^{2}\left( |C_{9}^{eff}|^{2}-|C_{10}|^{2}\right)
\left[ (F_{1}^{2}+F_{2}^{2})\ \hat{t}+2F_{1}F_{2}\sqrt{r}\right] \,.
\label{rate}
\end{eqnarray}
Here the form factors and Wilson coefficients in Eq. (\ref{rate}) depend on
the $\Lambda$ energy ($E_{\Lambda }$) and the scale of $\mu$. The ranges of $%
\hat{t}$ and $\hat{s}$ are as follows:
\begin{eqnarray}
\sqrt{r} &\leq &\hat{t}\leq \frac{1}{2}\left( 1+r-4\hat{m}_{l}^{2}\right) ,
\nonumber \\
4\hat{m}_{l}^{2} &\leq &\hat{s}\leq \left( 1-\sqrt{r}\right) ^{2}\,.
\end{eqnarray}
We note that our result for the differential decay rate in Eq. (\ref
{diffrate}) is consistent with that given in Refs. \cite{Lb1,Lb2} when one
takes the limit of massless lepton.

The longitudinal, normal and transverse $\Lambda $ polarization asymmetries
in Eq. (\ref{M1}) can be defined by
\begin{eqnarray}
P_{i}\left( \hat{t}\right)& =&\frac{d\Gamma \left( \hat{e}_{i}\cdot \hat{\xi}%
=1\right) -d\Gamma \left( \hat{e}_{i}\cdot \hat{\xi}=-1\right) }{d\Gamma
\left( \hat{e}_{i}\cdot \hat{\xi}=1\right) +d\Gamma \left( \hat{e}_{i}\cdot
\hat{\xi}=-1\right) }\,.  \label{asy}
\end{eqnarray}
 From Eqs. (\ref{diffrate}) and (\ref{asy}), we obtain the polarizations of $%
P_L$ and $P_N$ to be
\begin{eqnarray}
P_L( \hat{t}) &=&\frac{\sqrt{t^{2}-r}}{\sqrt{r}\rho _{0}\left( \hat{t}%
\right) }D_{L}
\end{eqnarray}
and
\begin{eqnarray}
P_{N}\left( \hat{t}\right) &=&\frac{-3}{2\rho _{0}\left( \hat{t}\right) }\pi
\sqrt{1-4\frac{\hat{m}_{l}^{2}}{\hat{s}}}\sqrt{\hat{s}\ }\left[ \left(
F_{1}^{2}+F_{2}^{2}\right) \sqrt{r}+2F_{1}F_{2}\ \hat{t}\right]  \nonumber \\
&&\times \left[ \mathop{\rm Re}C_{9}^{eff}C_{10}^{*}\left( 1-\hat{t}\right)
+2\hat{m}_{b}\mathop{\rm Re}C_{10}C_{7}^{*}\right] \,,  \label{PN}
\end{eqnarray}
respectively, where $D_{L}=L_{1}+L_{2}+L_{3}+L_{4}$ with
\begin{eqnarray}
L_{1} &=&-4{\frac{\hat{m}_b^2}{\widehat{s}}} \left(1-2\frac{\hat{m}_l^2}{%
\hat{s}}\right) |C_{7}|^{2}\sqrt{r} \left\{ - \left( 1-4\frac{\hat{m}_l^2}{%
\hat{s}} \right) \left( F_1^2-F_2^2\right) \ \hat{s} \right.  \nonumber \\
&& \left. +4\left( 1-\frac{\hat{m}_l^2}{\hat{s}}\right)
\left(F_1^2-F_2^2+2F_2^{2}\ \hat{t}+2F_{1}F_{2}\sqrt{r}\right) \left(
1-t\right) \right\}  \nonumber \\
&& +8\frac{\hat{m}_l^2\hat{m}_b^2}{\hat{s}}|C_{7}|^{2}\sqrt{r} \left\{
\left(F_1^2+F_2^2\right) \left( 1-10\frac{1-\hat{t}}{\hat{s}}\right)
+3\left( 1-2\frac{\hat{m}_l^2}{\hat{s}}\right)
\left(F_{1}^{2}-F_{2}^{2}\right) \right.  \nonumber \\
&& \left. -2\left( 1-4\frac{\hat{m}_l^2}{\hat{s}}\right) F_{2}^{2}+4\left(
5-2\frac{\hat{m}_l^2}{\hat{s}}\right) \left(\frac{1-\hat{t}}{\hat{s}}\right)
\left( F_2^2\left( 1-t\right) -F_1F_2\sqrt{r}\right) \right\} ,  \nonumber \\
L_{2} &=&-12\ \hat{m}_{b}\mathop{\rm Re}C_{9}^{eff}C_{7}^{*}\left( 1+2\frac{%
\hat{m}_{l}^{2}}{\hat{s}}\right) \sqrt{r}\left[ \left(
F_{1}^{2}-F_{2}^{2}\right) +2\ \hat{t}\ F_{2}^{2}+2\ \sqrt{r}\
F_{1}F_{2}\right] ,  \nonumber \\
L_{3} &=&-\left( |C_{9}^{eff}|^{2}+|C_{10}|^{2}\right) \sqrt{r}\left\{
\left( 1-4\frac{\hat{m}_{l}^{2}}{\hat{s}}\right) (F_{1}^{2}-F_{2}^{2})\ \
\hat{s}+2\ \left( 1+2\frac{\hat{m}_{l}^{2}}{\hat{s}}\right) \left( 1-\hat{t}%
\right) \times \right.  \nonumber \\
&&\left. \left[ (F_{1}^{2}-F_{2}^{2}+2\ \hat{t}\ F_{2}^{2}+2F_{1}F_{2}\
\sqrt{r}\right] \right\} ,  \nonumber \\
L_{4} &=&-6\hat{m}_{l}^{2}\left( |C_{9}^{eff}|^{2}-|C_{10}|^{2}\right)
(F_{1}^{2}-F_{2}^{2})\ \sqrt{r}\,.  \label{L}
\end{eqnarray}
For the T-odd transverse $\Lambda$ polarization, we have that
\begin{eqnarray}
P_{T} &\sim &m_{s} \mathop{\rm Im} C_{10}C_{7}^{*}\,.  \label{Pt}
\end{eqnarray}
It is clear that $P_T$ is zero in the standard model since there is no phase
in $C_{10}C_7^*$. We remark that even there is a phase in a theory of the
standard model like, due to the suppression of $m_s$, $P_T$ is expected to
be small. However, a possible CP violating right-handed interaction could
induce a sizable $P_T$ \cite{chen1}. 
Therefore, observing $P_T$ could indicate new physics beyond the standard
model.

It is interesting to point out that we can also discuss $\Lambda_b%
\rightarrow \Lambda \ \bar{\nu}\ \nu $ by taking the limits of
\begin{eqnarray}
m_l\rightarrow 0,\ C_7\rightarrow 0,\ C_9^{eff}\rightarrow \frac{X(x_t)}{%
\sin ^2\theta _W},\ C_{10}\rightarrow -\frac{X(x_t)}{\sin ^2\theta _W}
\end{eqnarray}
in Eqs. (\ref{Dr0})-(\ref{Pt}), where $X(x_t) =0.65x_{t}^{0.575}$ \cite
{Buras} and $x_t=m_t^2/M_W^2$. Explicitly, we have
\begin{eqnarray}
d\Gamma \left( \Lambda _{b}\rightarrow \Lambda \ \bar{\nu}\ \nu \right) &=&%
\frac{1}{2}d\Gamma ^{0}\left( \Lambda _{b}\rightarrow \Lambda \ \bar{\nu}\
\nu \right) \left[ 1+\vec{P}^{\nu \nu }\cdot \hat{\xi}\right] ,  \nonumber \\
d\Gamma ^{0}\left( \Lambda _{b}\rightarrow \Lambda \ \bar{\nu}\ \nu \right)
&=&3\frac{G_{F}^{2}\alpha _{em}^{2}\lambda _{t}^{2}}{192\pi ^{5}}M_{\Lambda
_{b}}^{5}\sqrt{\hat{t}^{2}-r}\ \rho ^{\nu \nu }\left( \hat{t}\right) d\cos
\theta _{\Lambda }d\hat{t}  \label{nunu}
\end{eqnarray}
where
\begin{eqnarray}
\rho ^{\nu \nu }\left( \hat{t}\right) &=&2\left( \frac{X\left( x_{t}\right)
}{\sin ^{2}\theta _{W}}\right) ^{2}\left\{ \left[ (F_{1}^{2}+F_{2}^{2})\
\hat{t}+2F_{1}F_{2}\sqrt{r}\right] \hat{s}+2\left( 1-\hat{t}\right) \right.
\nonumber \\
&&\left. \times \left[ \left( \hat{t}-r\right)
(F_{1}^{2}-F_{2}^{2})+2F_{2}(F_{1}\sqrt{r}+F_{2}\hat{t})\left( 1-\hat{t}%
\right) \right] \right\}  \label{rnunu}
\end{eqnarray}
and
\begin{eqnarray}
P_{L}^{\nu \nu }\left( \hat{t}\right) &=&-2\frac{\sqrt{t^{2}-r}}{\rho ^{\nu
\nu }\left( \hat{t}\right) }\left( \frac{X\left( x_{t}\right) }{\sin
^{2}\theta _{W}}\right) ^{2}\left\{ (F_{1}^{2}-F_{2}^{2})\ \hat{s}+2\ \left(
1-t\right) \times \right.  \nonumber \\
&&\left. \left[ (F_{1}^{2}-F_{2}^{2}+2\ \hat{t}\ F_{2}^{2}+2F_{1}F_{2}\
\sqrt{r}\right] \right\}\,.  \label{nnunu}
\end{eqnarray}
Here we have only listed the longitudinal polarization of $\Lambda$ because
the momentum of the neutrino cannot be measured experimentally.

\section{Numerical Analysis}

In order to analyze the decay rate and polarization asymmetries, we use the
Wilson coefficients at the scale $\mu \approx m_{b}$ as stated in Sec. 2.
The other parameters used in our numerical calculations are listed in Table
1.

\begin{table}[h]
\caption{ Input parameters used in our numerical calculations.}
\begin{center}
\begin{tabular}{|lll|}
\hline
$M_{\Lambda _{b}}$ & $5.64$ & GeV \\
$M_{\Lambda }$ & $1.116$ & GeV \\
$m_{t}$ & $165$ & GeV \\
$m_{b}$ & $4.8$ & GeV \\
$m_{\tau }$ & $1.777$ & GeV \\
$m_{\mu }$ & $1.05$ & GeV \\
$m_{c}$ & $1.4$ & GeV \\
$\alpha _{em}$ & $1/129$ &  \\
$\tau _{\Lambda _{b}}$ & $1.8848\times 10^{12}$ & GeV$^{-1}$ \\
$V_{tb}V_{ts}^{*}$ & $0.04$ &  \\ \hline
\end{tabular}
\end{center}
\end{table}

\noindent As to the $\Lambda _{b}\rightarrow \Lambda $ transition form
factors, we adopt the results and input parameters given in Ref. \cite{Lb1},
in which the QCD sum rule approach based on the framework of HQET was used.
However, there is a undetermined parameter, Borel parameter (M), in the
approach, which is introduced to suppress the contribution from the higher
excited and continuum states. According to the analysis of Ref. \cite{Lb1},
it could be $1.5\ GeV\leq M\leq 1.9\ GeV$. For simplicity, we will take $%
M=1.7\ GeV$ in our numerical analysis. As a comparison, we will also present
the results with the dipole form assumption \cite{MR}.

\subsection{Decay Rates and Polarizations of $\Lambda $}

 From Eqs. (\ref{diffrate}) and ({\ref{rate}), by integrating the
whole range of $\Lambda $ energy and setting phenomenological
factor $\kappa =-1/\left( 3C_{1}+C_{2}\right) $, the branching
ratios of the dilepton decays are summarized in Table 2 and the
distributions of the differential decay rates are shown in
Figures 1 and 2 for $\Lambda _{b}\to \Lambda \mu ^{+}\mu ^{-}$
and $\Lambda _{b}\to \Lambda \tau ^{+}\tau ^{-}$, respectively.
Here we have also illustrated the results from the pole model
\cite{MR}. The form factors with the dipole forms in the model
are given by
\begin{equation}
F_{1,2}\left( p_{\Lambda }\cdot v\right) =N_{1,2}\left( \frac{\Lambda _{QCD}%
}{\Lambda _{QCD}+p_{\Lambda }\cdot v}\right) ^{2}  \label{dipole}
\end{equation}
where $p_{\Lambda }\cdot v=E_{\Lambda }$ and $\Lambda _{QCD}$ is chosen to
be around $200$ $MeV$. From Eq. (\ref{dipole}),
one obtains that $R=F_{2}/F_{1}=N_{2}/N_{1}\sim -0.25$ \cite{MR,CLEO2}. In
terms of HQET the form factors of $\Lambda _{b}\rightarrow \Lambda $ should
be the same as that of $\Lambda _{c}\rightarrow \Lambda $ at the maximal
momentum transfer. Therefore, by using the measured branching ratio of $%
\Lambda _{c}\rightarrow \Lambda l\nu $, we extract that $|N_{1}|\sim 52.32$
with the same dipole forms. }

\begin{table}[h]
\caption{Decay branching ratios (Br) based on the form factors from the QCD
sum rule approach and the dipole model, respectively}
\begin{center}
\begin{tabular}{|l|l|l|l|l|l|}
\hline
Model & Decay Br & $\Lambda_b\to\Lambda\nu\bar{\nu}$ & $\Lambda_b\to \Lambda
e^+e^-$ & $\Lambda_{b}\to \Lambda \mu^{+}\mu^{-}$ & $\Lambda _{b}\to \Lambda
\tau^+\tau^-$ \\ \hline\hline
QCD & without LD & \multicolumn{1}{|c|}{$1.6\times 10^{-5}$} &
\multicolumn{1}{|c|}{$2.3\times 10^{-6}$} & \multicolumn{1}{|c|}{$2.1\times
10^{-6}$} & \multicolumn{1}{|c|}{$1.8\times 10^{-7}$} \\ \cline{2-6}
Sum Rule & with LD & \multicolumn{1}{|c|}{$--$} & \multicolumn{1}{|c|}{$%
5.3\times 10^{-5}$} & \multicolumn{1}{|c|}{$5.3\times 10^{-5}$} &
\multicolumn{1}{|c|}{$1.1\times 10^{-5}$} \\ \hline
Pole Model & without LD & \multicolumn{1}{|c|}{$9.2\times 10^{-6}$} &
\multicolumn{1}{|c|}{$1.2\times 10^{-6}$} & \multicolumn{1}{|c|}{$1.2\times
10^{-6}$} & \multicolumn{1}{|c|}{$2.6\times 10^{-7}$} \\ \cline{2-6}
\multicolumn{1}{|l|}{} & with LD & \multicolumn{1}{|c|}{$--$} &
\multicolumn{1}{|c|}{$3.6\times 10^{-5}$} & \multicolumn{1}{|c|}{$3.6\times
10^{-5}$} & \multicolumn{1}{|c|}{$9.0\times 10^{-6}$} \\ \hline
\end{tabular}
\end{center}
\end{table}


 From Table 2, we find that the branching ratios with including LD
contributions are about $1-2$ orders of magnitude larger than
that without LD ones and the results from the pole model are
close to those from the QCD sum rule.

If it is not mentioned, we shall use the form factors from the QCD sum rule
approach in the rest of our numerical analysis.

To estimate the contributions to the decay branching ratios by excluding the
resonances of $J/\psi $ and $\psi^{\prime}$, we choose five separate regions
in terms of the masses of 
$J/\psi $ and $\psi ^{\prime }$, and they are given as follows
\begin{eqnarray}
I &:&\ M_{\Lambda}\leq E_{\Lambda }\leq E|_{\max }-\delta _{\psi ^{\prime
}}^{1}\,,  \nonumber \\
II &:&\ E|_{\max }-\delta _{\psi ^{\prime }}^{1}\leq E_{\Lambda }\leq
E|_{\max }-\delta _{\psi ^{\prime }}^{2}\,,  \nonumber \\
III&:&\ E|_{\max }-\delta _{\psi ^{\prime }}^{2}\leq E_{\Lambda}\leq
E|_{\max }-\delta _{J/\psi }^{1}\,,  \nonumber \\
IV&:&\ E|_{\max}-\delta _{J/\psi}^1\leq E_{\Lambda}\leq E|_{\max} -\delta
_{J/\psi}^2,  \nonumber \\
V&:& E|_{\max}-\delta_{J/\psi }^{2}\leq E_{\Lambda }\leq E|_{\max }\,,
\end{eqnarray}
where
\[
\begin{array}{ll}
E|_{\max }=M_{\Lambda _{b}}\left( 1+r-4\hat{m}_{l}^{2}\right) /2, &  \\
\delta _{\psi ^{\prime }}^{1}=\left( M_{\psi ^{\prime }}+\sqrt{\sqrt{2}%
M_{\psi ^{\prime }}\Gamma _{\psi ^{\prime }}}\right) ^{2}/2M_{\Lambda _{b}},
& \delta _{\psi ^{\prime }}^{2}=\left( M_{\psi ^{\prime }}-\sqrt{\sqrt{2}%
M_{\psi ^{\prime }}\Gamma _{\psi ^{\prime }}}\right) ^{2}/2M_{\Lambda _{b}},
\\
\delta _{J/\psi }^{1}=\left( M_{J/\psi }+\sqrt{\sqrt{2}M_{J/\psi }\Gamma
_{J/\psi }}\right) ^{2}/2M_{\Lambda _{b}}, & \delta _{J/\psi }^{2}=\left(
M_{J/\psi }-\sqrt{\sqrt{2}M_{J/\psi }\Gamma _{J/\psi }}\right)
^{2}/2M_{\Lambda _{b}}.
\end{array}
\]
The factor of $\sqrt{2}$ in $\delta _{V}^{i}$ is a typical value and one may
take a larger value to reduce the LD contributions in the regions of $I$ and
$V$. The estimations of the decay branching ratios in the different regions
are listed in Table 3. From the table, We find that the RE in region I is
about $20\%$ for the $e^{+}e^{-}$ and $\mu ^{+}\mu ^{-}$ modes and $25\%$
for $\tau ^{+}\tau^{-}$. The larger RE for the $\tau $ pair arises from $%
\Gamma _{4}$ in Eq. (\ref{rate}), which is proportional to the lepton mass.
Moreover, this term also yields different distributions between the electron
(or muon) and tau modes in region I when a large deviation from $%
(|C_9(m_b)|-|C_{10}|)$ appears. Therefore, studying the region with lower RE
could distinguish the SD Wilson coefficients from the standard model.

\begin{table}[h]
\caption{ Decay branching ratios for QCD sume rule (SR) and Pole
model (PM) with and without LD  in different regions of $\Lambda $
energy with $\kappa =-1/\left( 3C_{1}+C_{2}\right) $. }

\begin{center}
\begin{tabular}{|c|cccccccccc|}
\hline &  &  &  & Br &  & ($\times 10^{-7}$) &  &  &  &  \\
\cline{2-11} Mode & I &  & \multicolumn{1}{|c}{II ($\times
10^{2}$)} &  & \multicolumn{1}{|c}{III} &  &
\multicolumn{1}{|c}{IV ($\times 10^{2}$)} &  &
\multicolumn{1}{|c}{V} & \multicolumn{1}{c|}{}
\\ \cline{2-11}
& SR & \multicolumn{1}{|c}{PM} & \multicolumn{1}{|c}{SR} &
\multicolumn{1}{|c}{PM} & \multicolumn{1}{|c}{SR} &
\multicolumn{1}{|c}{ PM} & \multicolumn{1}{|c}{SR} &
\multicolumn{1}{|c}{PM} & \multicolumn{1}{|c}{SR} &
\multicolumn{1}{|c|}{PM} \\ \hline \multicolumn{1}{|r|}{$ee$,$\ $
LD} & $2.7$ & \multicolumn{1}{|c}{$4.0$} &
\multicolumn{1}{|c}{$2.7$} & \multicolumn{1}{|c}{$2.3$} &
\multicolumn{1}{|c}{$3.9$} & \multicolumn{1}{|c}{$2.6$} &
\multicolumn{1}{|c}{$2.4$} & \multicolumn{1}{|c}{$1.2$} &
\multicolumn{1}{|c}{$19.6$} & \multicolumn{1}{|c|}{$6.6$} \\
\cline{2-11} \multicolumn{1}{|r|}{NLD} & $3.4$ &
\multicolumn{1}{|c}{$4.9$} & \multicolumn{1}{|c}{$0.005$} &
\multicolumn{1}{|c}{$0.004$} & \multicolumn{1}{|c}{$3.8$} &
\multicolumn{1}{|c}{$2.5$} & \multicolumn{1}{|c}{$0.003$} &
\multicolumn{1}{|c}{$0.001$} & \multicolumn{1}{|c}{$14.6$} &
\multicolumn{1}{|c|}{$4.4$} \\ \hline \multicolumn{1}{|r|}{$\ \mu
\mu ,$ LD } & $2.7$ & \multicolumn{1}{|c}{$4.0$} &
\multicolumn{1}{|c}{$2.7$} & \multicolumn{1}{|c}{$2.3$} &
\multicolumn{1}{|c}{$3.9$} & \multicolumn{1}{|c}{$2.6$} &
\multicolumn{1}{|c}{$2.4$} & \multicolumn{1}{|c}{$1.2$} &
\multicolumn{1}{|c}{$17.9$} & \multicolumn{1}{|c|}{$6.2$} \\
\cline{2-11} \multicolumn{1}{|r|}{$\ $ NLD} & $3.4$ &
\multicolumn{1}{|c}{$4.9$} & \multicolumn{1}{|c}{$0.005$} &
\multicolumn{1}{|c}{$0.004$} & \multicolumn{1}{|c}{$3.8$} &
\multicolumn{1}{|c}{$2.5$} & \multicolumn{1}{|c}{$0.003$} &
\multicolumn{1}{|c}{$0.001$} & \multicolumn{1}{|c}{$12.9$} &
\multicolumn{1}{|c|}{$4.0$} \\ \hline \multicolumn{1}{|r|}{$\ \tau
\tau $, LD} & $1.2$ & \multicolumn{1}{|c}{$1.9$} &
\multicolumn{1}{|c}{$1.0$} & \multicolumn{1}{|c}{$0.9$} &
\multicolumn{1}{|c}{$0.2$} & \multicolumn{1}{|c}{$0.1$} &
\multicolumn{1}{|c}{$--$} & \multicolumn{1}{|c}{$--$} & \multicolumn{1}{|c}{$%
--$} & \multicolumn{1}{|c|}{$--$} \\ \cline{2-11}
\multicolumn{1}{|r|}{$\ $NLD} & $1.6$ & \multicolumn{1}{|c}{$2.4$}
& \multicolumn{1}{|c}{$0.001$} & \multicolumn{1}{|c}{$0.001$} &
\multicolumn{1}{|c}{$1.1$} & \multicolumn{1}{|c}{$0.08$} &
\multicolumn{1}{|c}{$--$} & \multicolumn{1}{|c}{$--$} & \multicolumn{1}{|c}{$%
--$} & \multicolumn{1}{|c|}{$--$} \\ \hline
\end{tabular}
\end{center}
\end{table}

As we can see from Eq. (\ref{effc9}), the LD effects have been absorbed into
the Wilson coefficient of $C_{9}^{eff}$ and they are parametrized in the
form of the phenomenological Breit-Wigner Ansatz. To compensate FA and VMD
approximation, one phenomenological factor $\kappa $ is also introduced. In
Table 4, we show the decay branching ratios by taking $\kappa=-3.5$ and $%
-1.9 $. It is easily seen that in the regions of $I$ and $V$ the differences
for the branching ratios with lower and higher $\kappa$ are between $5\% -
16\%$. 
This tells us that, as expected, the uncertainty from the LD effect is small
outside the resonance region.

\begin{table}[h]
\caption{Decay branching ratios in the whole range of $\Lambda $ energy
including LD with two different values of $\kappa $.}
\begin{center}
\medskip
\begin{tabular}{|c|ccccc|}
\hline
&  & Br & ($\times 10^{-7}$) &  &  \\ \cline{2-6}
Decay Mode & I & \multicolumn{1}{|c}{II} & \multicolumn{1}{|c}{III} &
\multicolumn{1}{|c}{IV} & \multicolumn{1}{|c|}{V} \\ \hline
\multicolumn{1}{|l|}{$\Lambda _{b}\rightarrow \Lambda \ e^{+}\ e^{-}$, $\
\kappa =-3.5$} & $2.6$ & \multicolumn{1}{|c}{$5.7\times 10^2$} &
\multicolumn{1}{|c}{$4.9$} & \multicolumn{1}{|c}{$5.0\times 10^2$} &
\multicolumn{1}{|c|}{$23.2$} \\ \hline
\multicolumn{1}{|r|}{$\kappa =-1.9$} & $2.8$ & \multicolumn{1}{|c}{$%
1.7\times 10^2$} & \multicolumn{1}{|c}{$3.7$} & \multicolumn{1}{|c}{$%
1.5\times 10^2$} & \multicolumn{1}{|c|}{$18.3$} \\ \hline
\multicolumn{1}{|l|}{$\Lambda _{b}\rightarrow \Lambda \ \mu ^{+}\ \mu ^{-}$,
$\kappa =-3.5$} & $2.6$ & \multicolumn{1}{|c}{$5.7\times 10^2$} &
\multicolumn{1}{|c}{$4.9$} & \multicolumn{1}{|c}{$5.0\times 10^2$} &
\multicolumn{1}{|c|}{$21.4$} \\ \hline
\multicolumn{1}{|r|}{$\kappa =-1.9$} & $2.8$ & \multicolumn{1}{|c}{$%
1.7\times 10^2$} & \multicolumn{1}{|c}{$3.7$} & \multicolumn{1}{|c}{$%
1.5\times 10^2$} & \multicolumn{1}{|c|}{$16.5$} \\ \hline
\multicolumn{1}{|l|}{$\Lambda _{b}\rightarrow \Lambda \ \tau ^{+}\ \tau ^{-}$%
, $\ \kappa =-3.5$} & $1.1$ & \multicolumn{1}{|c}{$2.2\times 10^2$} &
\multicolumn{1}{|c}{$0.2$} & \multicolumn{1}{|c}{$--$} &
\multicolumn{1}{|c|}{$--$} \\ \hline
\multicolumn{1}{|r|}{$\kappa =-1.9$} & $1.2$ & \multicolumn{1}{|c}{$%
0.7\times 10^2$} & \multicolumn{1}{|c}{$0.2$} & \multicolumn{1}{|c}{$--$} &
\multicolumn{1}{|c|}{$--$} \\ \hline
\end{tabular}
\end{center}
\end{table}

In order to study how the effects arising from new physics beyond the
standard model will affect the baryonic dilepton decays, we consider cases
where the Wilson coefficients are different from those in the standard
model. The results for the distributions of the differential branching rates
are shown in Figures $3-6$.

We now discuss our results as follows:

$\bullet$ According to the results in Table 3 and Figures 1 and 2,
we clearly see that outside the resonant regions the uncertainties
arising from the QCD models are larger than that from the LD effects. %

$\bullet$ We first compare our results in baryon decays with
those in the B meson dilepton ones of $B\rightarrow
K^{*}l^{+}l^{-}$ \cite {DTP,LMS,AMM,OT,KS,Geng1}. In the meson
decays, the pole of $\hat{s}$ is
related to $\left| \hat{m}_{b}C_{7}/\hat{s}\right| ^{2}$ and $\hat{m}%
_{b}C_{7}/\hat{s}$, respectively, and thus with the requirement $\hat{s}\geq
4\hat{m}_{l}$ from the phase space, the processes of $B\rightarrow K^{*}\mu
^{+}\mu ^{-}$ and $B\rightarrow K^{*}e^{+}e^{-}$ have very different decay
rates. However, for the decays of $\Lambda _{b}\rightarrow\Lambda l^{+}l^{-}$%
, the associated terms are proportional to $\left| \hat{m}_{l}\hat{m}%
_{b}C_{7}\right| ^{2}/\hat{s}^{2}$ and $\left| \hat{m}_{b}C_{7}\right| ^{2}/%
\hat{s}$. Clearly, due to the mass suppression for the light lepton, the
main pole dependence is $\sim \left| \hat{m}_{b}C_{7}\right| ^{2}/\hat{s}$
so that the rate difference between $\Lambda_b\to \Lambda \mu^{+}\mu^{-}$
and $\Lambda_b\to \Lambda e^{+}e^{-}$ is small.

$\bullet$ The differential decay rates of $\Lambda _{b}\rightarrow \Lambda
l^{+}l^{-}$ are sensitive to the signs of $C_9$ and $C_7$. Although $%
C_{7}\ll C_{9}$ and $C_{10}$, there exists an enhanced factor of $12%
\mathop{\rm Re}C_{9}C_{7}^{*}\sim 15.3$ in $\Gamma _{2}$ of Eq. (\ref{rate}%
). When the sign of $C_{7}$ is opposite to that in the standard model, there
is a deviation of 50\% in branching ratio for neglecting RE. Thus, the
contribution from electromagnetic part cannot be neglected. As changing the
sign of $C_{9}$ to be opposite to that in the standard model, only a
deviation of 27\% on the branching ratios occurs. However, from Figures 3
and 5, we see that the distributions are different from each others.

$\bullet$ From Eq. (\ref{rate}), we find that the differential decay rates
cannot have the information in the sign of $C_{10}$ since they are always
related to $\left| C_{10}\right| ^{2}$.

$\bullet$ From Figures 4 and 6, we find that with $C_{10}=2C_{9}|_{SM}$ the
distribution for the differential decay rate of the $\tau ^{+}\tau ^{-}$
mode is higher than that with $C_{9}=2C_{10}|_{SM}$ in region I but it is
reversed in that of the $\mu^{+}\mu ^{-}$ distribution. The origin of this
difference is from the $\Gamma _{4}$ in Eq. ( \ref{rate}) which is
proportional to $6\hat{m}_{l}^{2}\left( \left| C_{9}\right| ^{2}-\left|
C_{10}\right| ^{2}\right) $. This effect can be neglected in the standard
model since $|C_{9}| \sim |C_{10}|$ and the light lepton modes as well.
Although $\hat{m}_{\tau }^{2}\sim 10\%$, this factor will become important.
if there is a large deviation between $C_{9}$ and $C_{10}$. 

$\bullet$ The decay width distributions for the longitudinal polarized $%
\Lambda$ with and without LD effects as the function of $\Lambda $ energy
are shown in Figures 7 and 8. Comparing the figures with the differential
decay branching rates in Figures 1 and 2, respectively, we find that the
distributions are very similar to each other except the opposite sign.

$\bullet$ Finally, as usual, from Eq. (\ref{diffrate}) we may also write the
partial decay rate as
\begin{eqnarray}
d\Gamma _{\Lambda _{b}}&=&\frac{1}{2}\Gamma _{0}\left( 1-\alpha _{\Lambda }\
{\rm \hat{p}\cdot \hat{s}} d\cos \theta _{\Lambda }\right)\,,
\end{eqnarray}
where $\Gamma _{0}$ is related to the decay width of $\Lambda
_{b}\rightarrow \Lambda \ l^{+}\ l^{-}$, \^{p} is the unit direction of $%
\Lambda $ momentum in the $\Lambda _{b}$ rest frame, and $\alpha _{\Lambda }$%
, called $\Lambda $ polarization, is defined by
\begin{eqnarray}
\alpha _{\Lambda }=\frac{\int_{r}^{t_{\max }}D_{L}(1-4 \hat{m}_l^2/\hat{s}%
)\left( \hat{t}^{2}-r\right) /\sqrt{r}d\hat{t}}{\int_{r}^{t_{\max }}\sqrt{1-4%
\hat{m}_l^2/\hat{s}}\sqrt{\hat{t}^{2}-r}\rho _{0}\left( \hat{t}\right) d\hat{%
t}},
\end{eqnarray}
where $t_{\max }=\left( 1+r-4\hat{m}_{l}^{2}\right) /2$. Numerically, we
find that the polarizations of $\Lambda $ in $\Lambda _{b}\rightarrow
\Lambda \ l^{+}\ l^{-}$ $(l=e,\ \mu ,\ \tau)$ decays are all unity, $\alpha
_{\Lambda }\approx 1$.


\subsection{Polarization asymmetries}

In this subsection we will discuss longitudinal and normal polarization
asymmetries and their implications and we will study the transverse
polarization elsewhere \cite{chen1} since it is zero in the standard model
as mentioned in Sec. 3. From Eq. (\ref{asy}), we show the the distributions
of $P_L$ and $P_N$ with respect to the dimensionless kinematic variable $%
\hat{t}$ in Figures $9-12$, respectively. From the figures, we find the
following interesting results:\newline

$\bullet$ The polarization asymmetries are insensitive to the LD effects.

$\bullet$ The values of $P_{L}$ are near unity except a narrow region with a
small $\Lambda $ momentum.

$\bullet$ $P_N$ approaches zero as the $\Lambda $ energy increases. This is
because the polarization is proportional to $\sqrt{\hat{s}}$ as shown in Eq.
(\ref{PN}).

$\bullet$ The values of $P_{L,N}$ from the QCD sum rule and the pole models
shown in the figures are close to each other. The results imply that both $%
P_L$ and $P_N$ are not very sensitive to the form factors. Therefore, one
would like to use $P_{L,N}$ to probe the short-distance (SD) physics due to
the smallness of the uncertainties from the strong interaction.\newline

We now discuss the sensitivity for the longitudinal polarization of $P_{L}$
to new physics. We first notice that by using different values of the Wilson
coefficients from the standard model, the polarizations do not change. The
reason is that the coefficients get canceled out between the denominator and
numerator in Eq. (\ref{asy}). However, in our derivation for the
differential decay rate, we have assumed the $V-A$ hadronic current and
neglected the contribution of left-handed electromagnetic moment since it is
proportional to the strange quark mass. If we include the interaction with
the right-handed current, the polarization will behave quit different from
that in the standard model, which can be understand easily by Eq. (26) of
Ref. \cite{Lb1} as $h_A\neq h_V$. Finally, we define the integrated
longitudinal and normal polarization asymmetries as
\begin{eqnarray}
\bar{P}_{L}&=&\int d\hat{t} P_L\,,  \nonumber \\
\bar{P}_{N}&=&\int d\hat{t} P_N\,.  \label{plpn}
\end{eqnarray}
In the standard model, we obtain that $\bar{P}_{L(N)}=-0.31\ ( 0.02) $ and $%
\bar{P}_{L(N) }=-0.12\ (0.01)$ for $\mu \mu $ and $\tau \tau $ modes,
respectively. If deviations from the standard model predictions for the
integrated polarization asymmetries are measured, it is clear that there
exit some kinds of new physics.

\section{Conclusions}

We have studied the rare baryonic exclusive decays of $\Lambda_b\to \Lambda
l^+ l^-\ (l=e,\mu,\tau )$ with polarized $\Lambda $. Under the approximation
of HQET, in the standard model we have derived the differential decay rates
and the polarization asymmetries of $\Lambda $ by including lepton mass
effects.

We have found that with the LD effects the decay branching ratios of $%
\Lambda_b\to \Lambda l^+ l^-\ (l=e,\mu,\tau )$ are $5.3\times 10^{-5}$, $%
5.3\times 10^{-5}$, and $1.1\times 10^{-5}$ from the QCD sum rule approach
and $1.2\times 10^{-5}$, $1.2\times 10^{-5}$, and $3.2\times 10^{-6}$ from
the the pole model, respectively. We have also estimated the decay branching
ratio of $\Lambda_b\to \Lambda \nu \bar{\nu}$ to be $1.6\times 10^{-5}$ and $%
3.3\times 10^{-6}$ in the two models, respectively. In physics beyond the
standard model, we have studied various cases of different Wilson
coefficients. We have shown that the decay rates as well as the
distributions can be very different from those in the standard model.

The integrated longitudinal $\Lambda$ polarizations are $-0.31$ and $-0.12$,
while that of the normal ones $0.02$ and $0.01$, for di-muon and tau modes,
respectively. The CP-odd transverse polarization of $\Lambda$ is zero in the
standard model but it is expected to be sizable in new physics such as the
CP violating theories with right-handed interactions. We have demonstrated
that the polarization asymmetries are insensitive to LD contributions but
sensitive to the right-handed couplings. It is clear that one could probe
new physics through measurements of the $\Lambda$ polarizations in the
decays of $\Lambda_b\to \Lambda l^+ l^-$. \newline

{\bf Acknowledgments}

This work was supported in part by the National Science Council of the
Republic of China under contract numbers NSC-89-2112-M-007-054 and
NSC-89-2112-M-006-033.

\newpage 

\newpage
\begin{figcap}
\item
The  differential decay branching ratio of $\Lambda_b\to\Lambda
\mu^+\mu^-$ as a function of $\Lambda$ energy. The solid and
dashed curves stand for the QCD sum rule and pole models,
respectively.

\item
Same as Figure 1 but for $\Lambda_b\to\Lambda \tau^+\tau^-$.

\item
The  differential decay branching ratio of $\Lambda_b\to\Lambda
\mu^+\mu^-$ as a function of $\Lambda$ energy with the Wilson
coefficients being different from those in the standard model. The
solid, dashed, dotted, long-dashed, and dash-dotted curves
represent the results of the standard model (SM), $C_{10}=0$,
$C_9=-C_9|_{SM}$, $C_7=-C_7|_{SM}$, and $C_{7}=0$, respectively.

\item
Same as Figure 3 but the
 dashed, dotted, long-dashed, and dash-dotted curves
are for $C_9=-2C_9|_{SM}$, $C_9=-2C_9|_{SM}$ and $C_7=0$,
$C_9=2C_9|_{SM}$, and $C_{10}=2C_{10}|_{SM}$, respectively.

\item
Same as Figure 3 but for $\Lambda_b\to\Lambda \tau^+\tau^-$.

\item
Same as Figure 4 but for $\Lambda_b\to\Lambda \tau^+\tau^-$.

\item
The decay width distribution of $\Lambda_b\to\Lambda \mu^+\mu^-$
for the longitudinal polarized $\Lambda$ as a function of
$\Lambda$ energy  with and without RE.

\item
Same as Figure 7 but for $\Lambda_b\to\Lambda \tau^+\tau^-$.

\item
The longitudinal polarization asymmetry of $\Lambda_b\to\Lambda
\mu^+\mu^-$ as a function of $E_{\Lambda}/M_{\Lambda}$. Legend is
the same as Figure 1.

\item
Same as Figure 9 but for $\Lambda_b\to\Lambda \tau^+\tau^-$.

\item
The normal polarization asymmetry of $\Lambda_b\to\Lambda
\mu^+\mu^-$ as a function of $E_{\Lambda}/M_{\Lambda}$. Legend is
the same as Figure 1.

\item
Same as Figure 11 but for $\Lambda_b\to\Lambda \tau^+\tau^-$.

\end{figcap}

\newpage

\begin{figure}[h]
\includegraphics{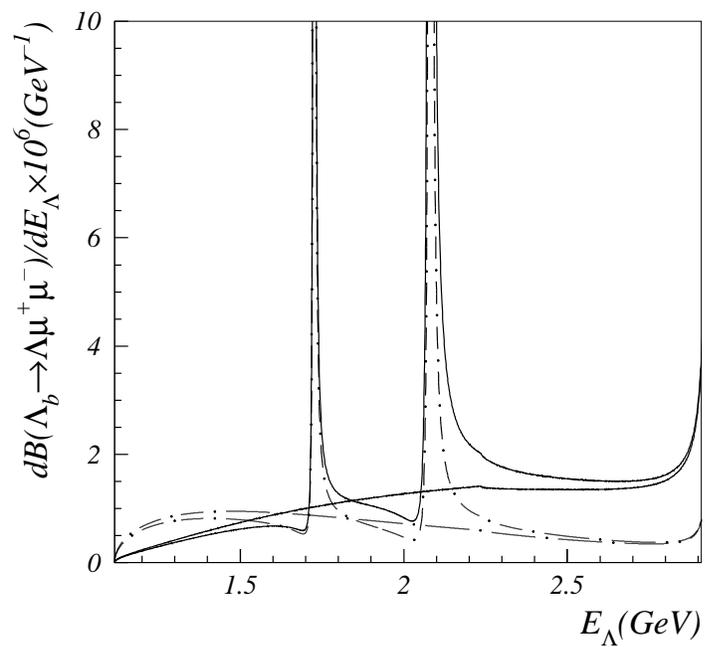} %
\vskip 15.5cm
\caption{ The differential decay branching ratio of $\Lambda_b\to\Lambda
\mu^+\mu^-$ as a function of $\Lambda$ energy. The solid and dashed curves
stand for the QCD sum rule and pole models, respectively. }
\end{figure}

\begin{figure}[h]
\includegraphics{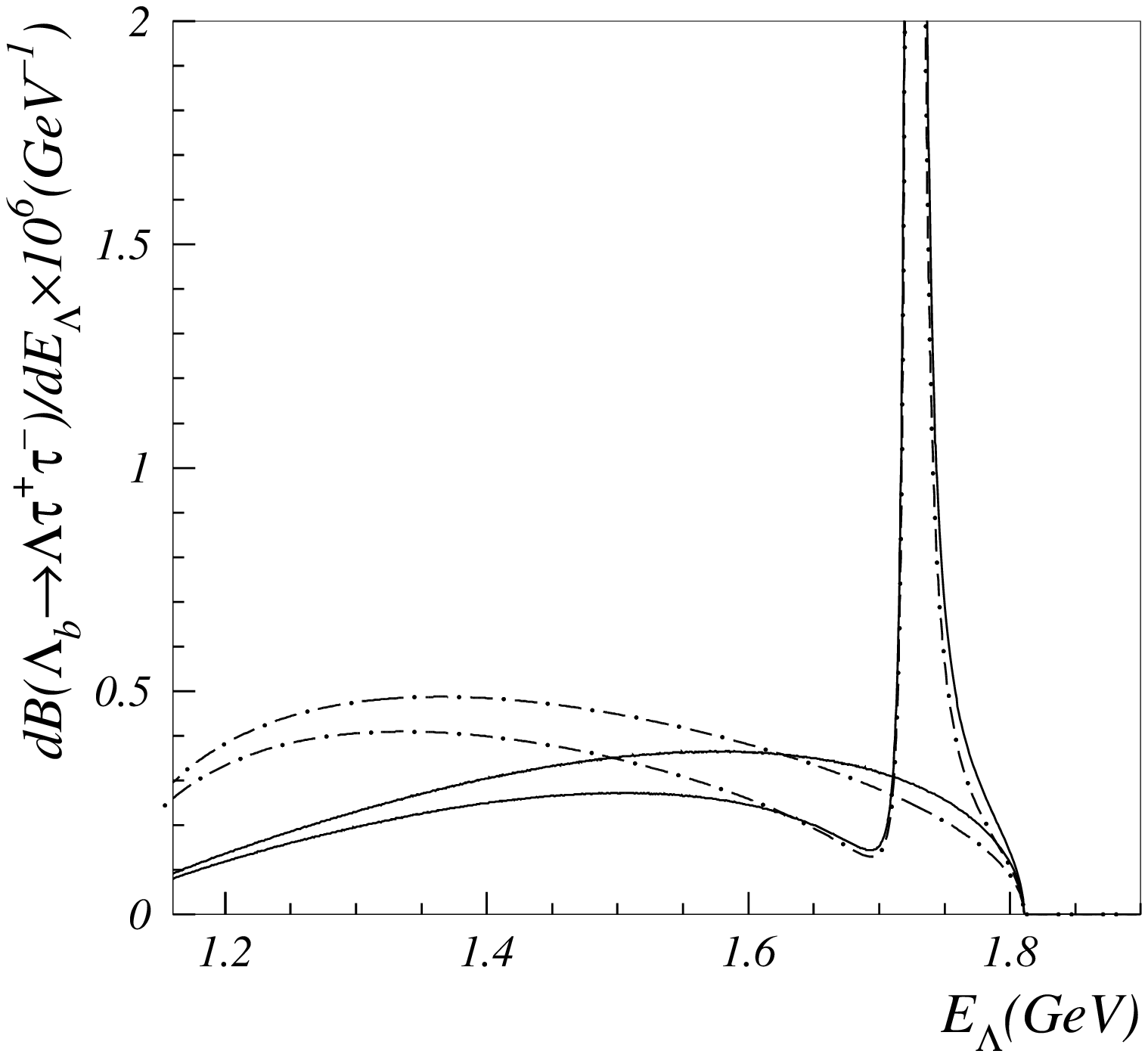} %
\vskip 11cm
\caption{Same as Figure 1 but for $\Lambda_b\to\Lambda \tau^+\tau^-$. }
\end{figure}

\begin{figure}[h]
\includegraphics{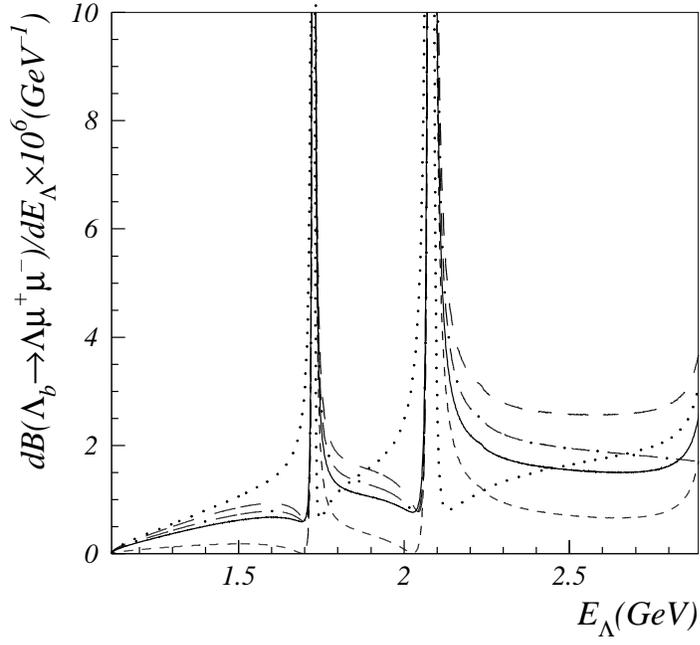} %
\vskip 11cm
\caption{The differential decay branching ratio of $\Lambda_b\to\Lambda
\mu^+\mu^-$ as a function of $\Lambda$ energy with the Wilson coefficients
being different from those in the standard model. The solid, dashed, dotted,
long-dashed, and dash-dotted curves represent the results of the standard
model (SM), $C_{10}=0$, $C_9=-C_9|_{SM}$, $C_7=-C_7|_{SM}$, and $C_{7}=0$,
respectively. }
\end{figure}

\begin{figure}[h]
\includegraphics{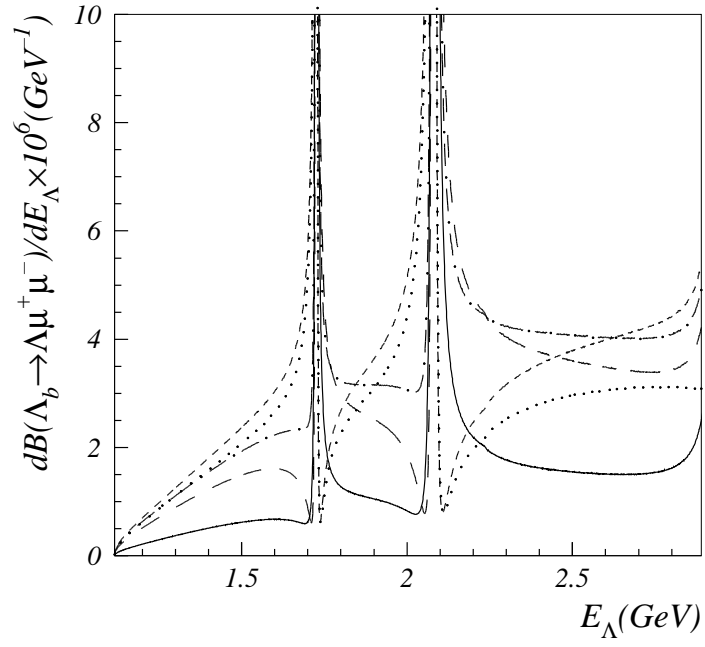} %
\vskip 11cm
\caption{Same as Figure 3 but the dashed, dotted, long-dashed, and
dash-dotted curves are for $C_9=-2C_9|_{SM}$, $C_9=-2C_9|_{SM}$ and $C_7=0$,
$C_9=2C_9|_{SM}$, and $C_{10}=2C_{10}|_{SM}$, respectively. }
\end{figure}

\begin{figure}[h]
\includegraphics{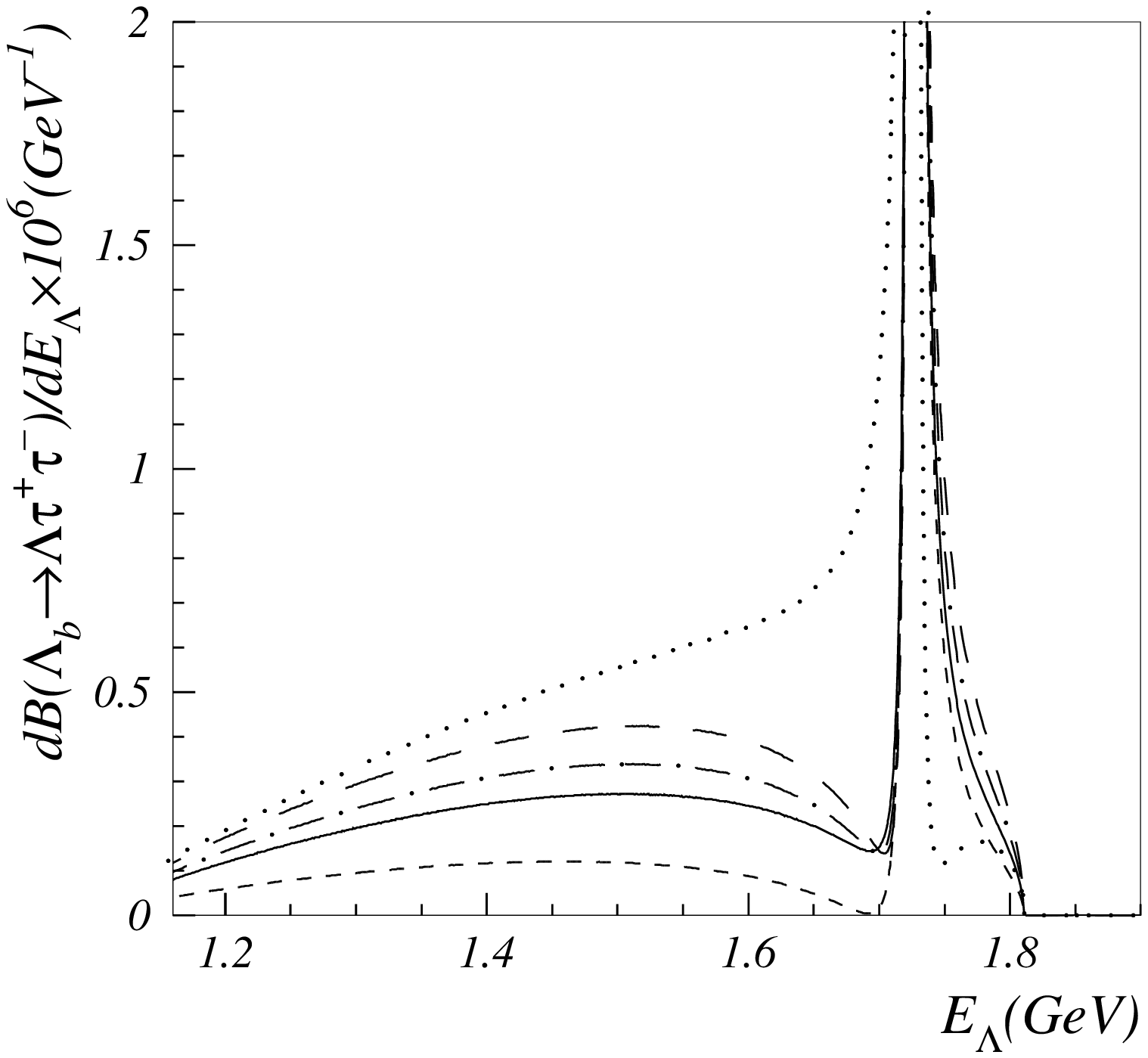} %
\vskip 12cm
\caption{Same as Figure 3 but for $\Lambda_b\to\Lambda \tau^+\tau^-$. }
\end{figure}

\begin{figure}[h]
\includegraphics{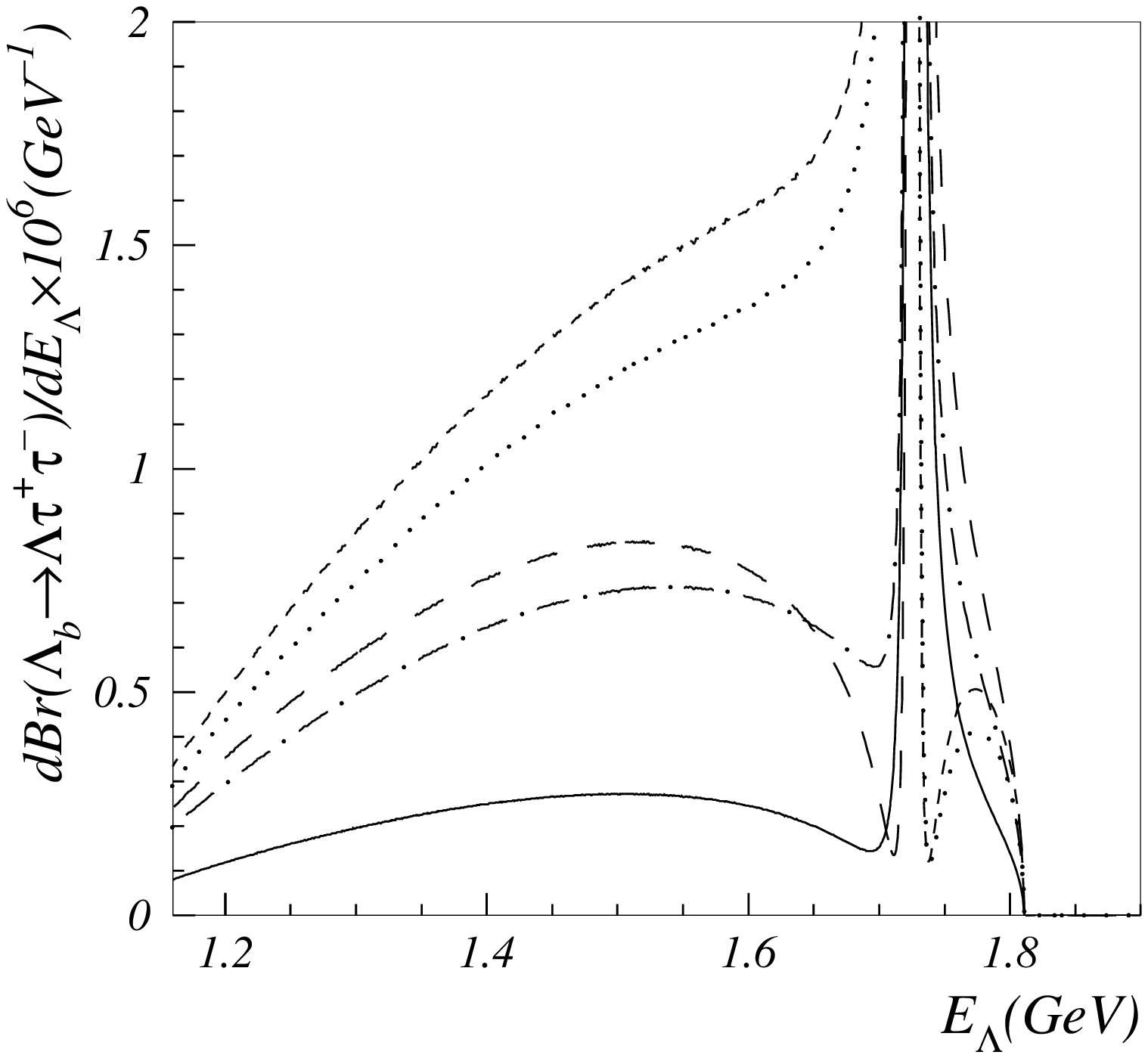} %
\vskip 12cm
\caption{Same as Figure 4 but for $\Lambda_b\to\Lambda \tau^+\tau^-$. }
\end{figure}

\begin{figure}[h]
\includegraphics{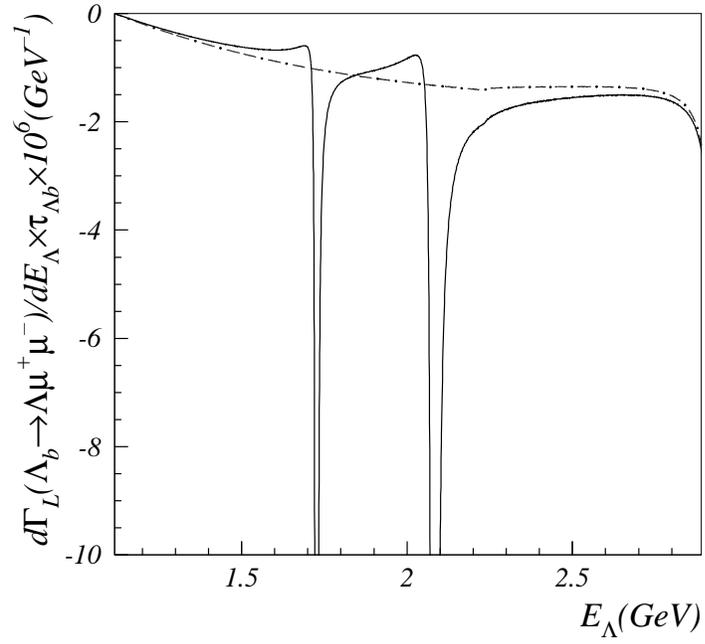} %
\vskip 11cm
\caption{The decay width distribution of $\Lambda_b\to\Lambda \mu^+\mu^-$
for the longitudinal polarized $\Lambda$ as a function of $\Lambda$ energy
with and without RE. }
\end{figure}

\begin{figure}[h]
\includegraphics{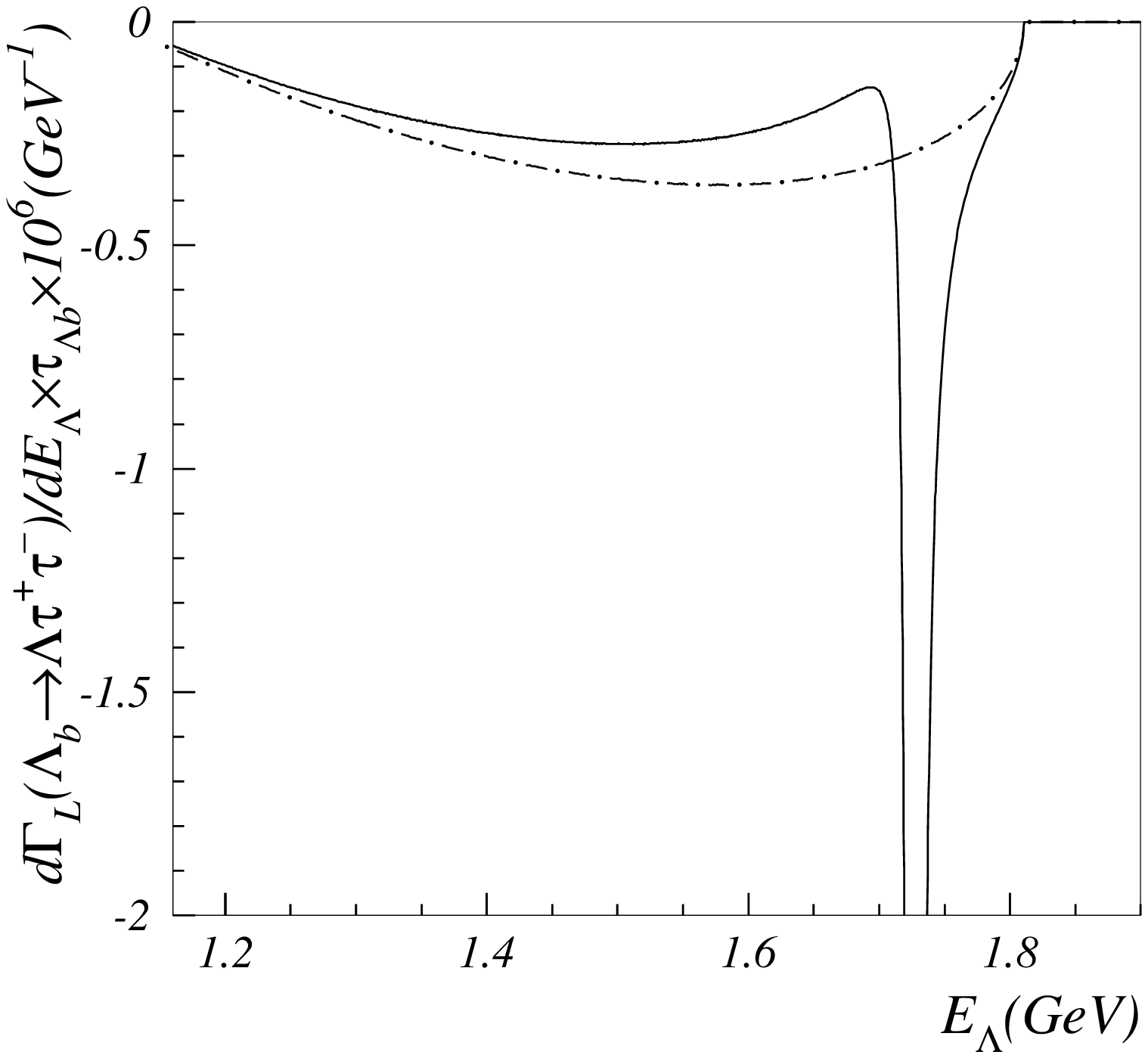} %
\vskip 11cm
\caption{Same as Figure 7 but for $\Lambda_b\to\Lambda \tau^+\tau^-$. }
\end{figure}

\begin{figure}[h]
\includegraphics{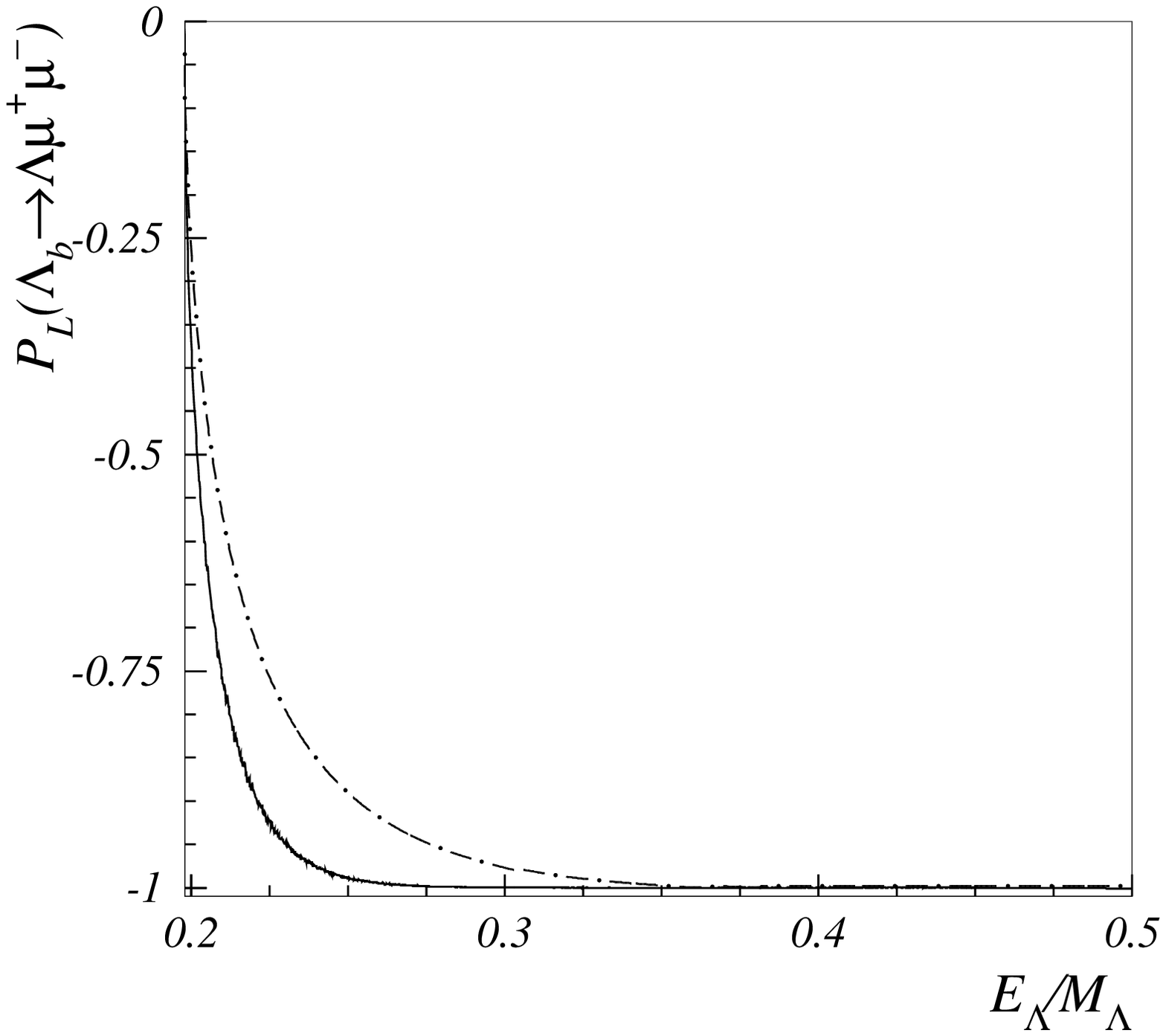} %
\vskip 11cm
\caption{The longitudinal polarization asymmetry of $\Lambda_b\to\Lambda
\mu^+\mu^-$ as function of $E_{\Lambda}/M_{\Lambda}$. Legend is the same as
Figure 1. }
\end{figure}

\begin{figure}[h]
\includegraphics{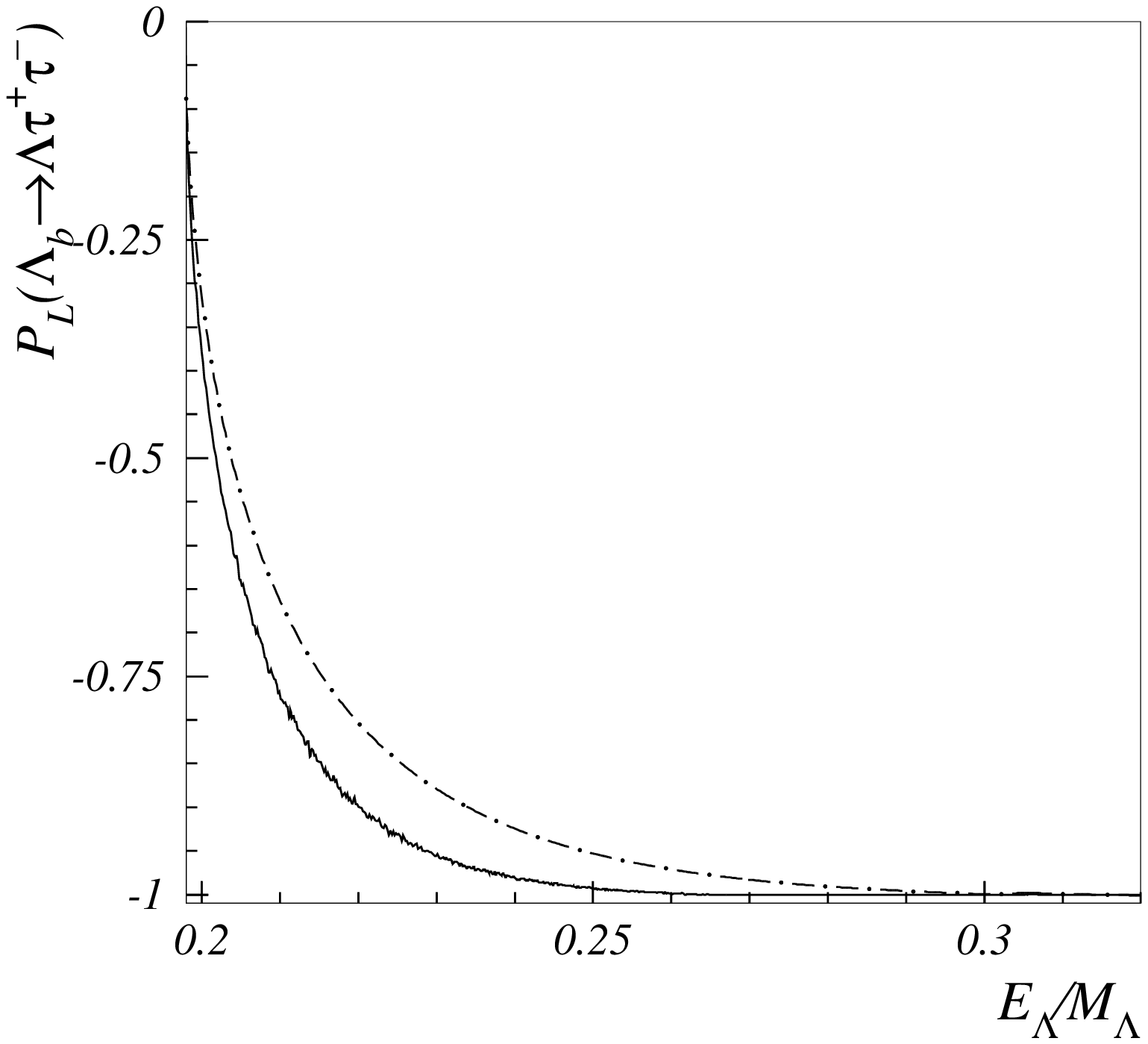}
\vskip 11cm
\caption{Same as Figure 9 but for $\Lambda_b\to\Lambda \tau^+\tau^-$. }
\end{figure}

\begin{figure}[h]
\includegraphics{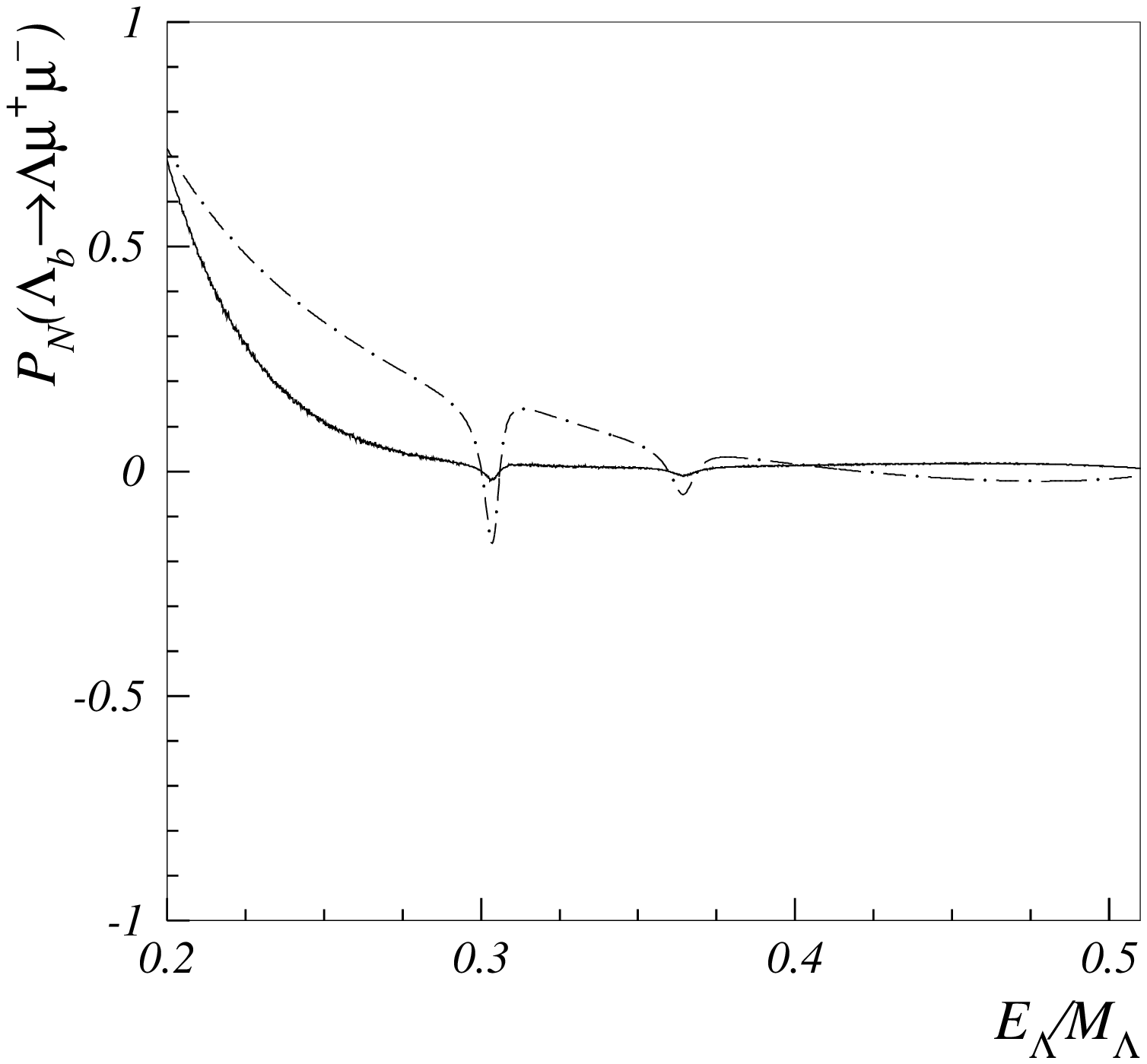}
\vskip 11cm
\caption{The normal polarization asymmetry of $\Lambda_b\to\Lambda
\mu^+\mu^- $ as function of $E_{\Lambda}/M_{\Lambda}$. Legend is the same as
Figure 1. }
\end{figure}

\begin{figure}[h]
\includegraphics{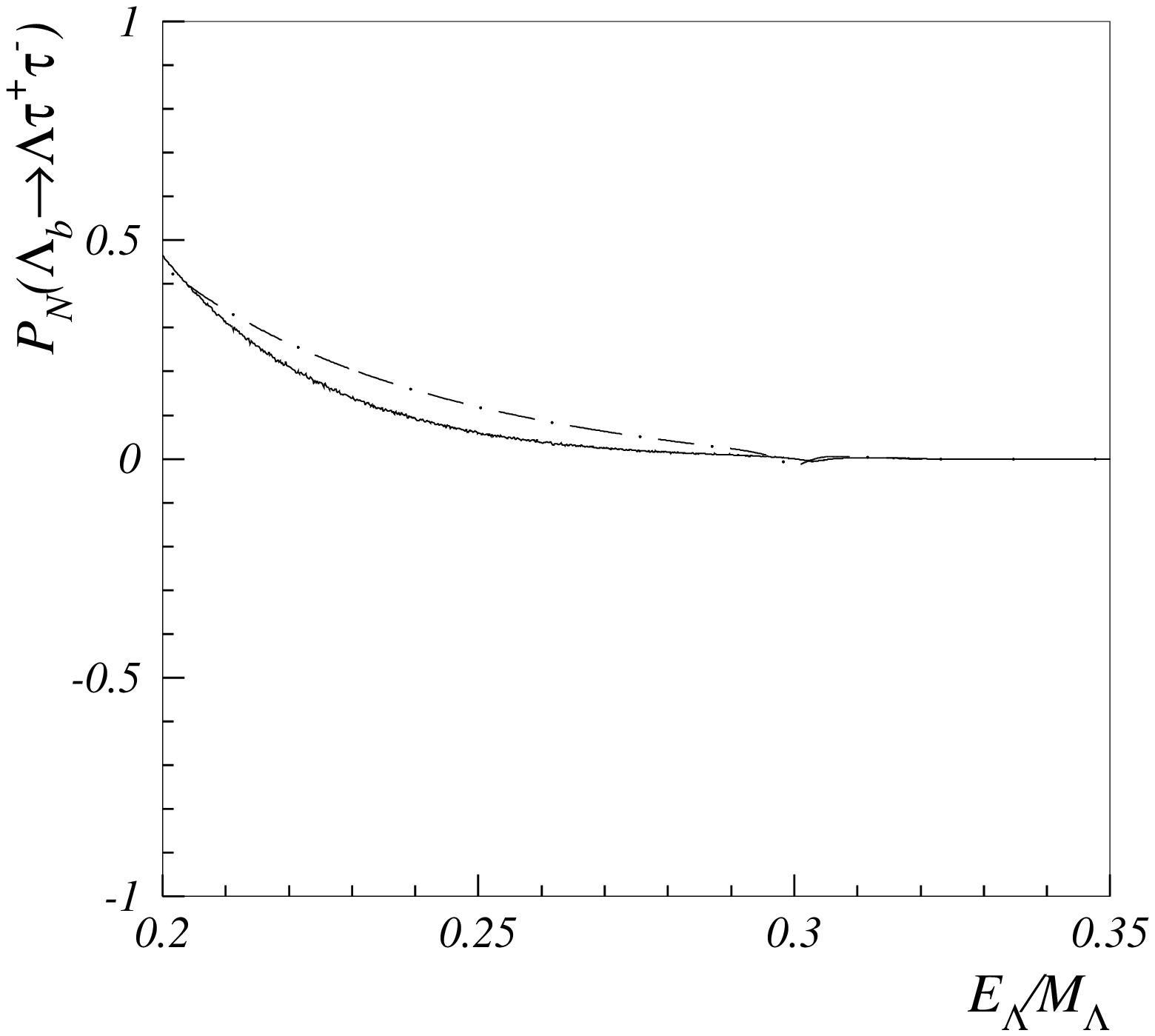}
\vskip 11cm
\caption{Same as Figure 11 but for $\Lambda_b\to\Lambda \tau^+\tau^-$. }
\end{figure}

\end{document}